\documentclass[twocolumn]{article}\usepackage{latexsym,amsmath,amssymb,amsfonts,amsthm,epsfig}
\newcommand\INDENT{\indent\indent}
\newcommand\ignore[1]{}

\newcommand\Tr{\mathop{\mathrm{Tr}}\nolimits}
\newcommand\Range{\mathop{\mathrm{Range}}\nolimits}
\newcommand\HH{\mathcal{H}}  
\newcommand\calS{\mathcal{S}}

\newcommand\ox[1]{{\otimes{#1}}} 
\newcommand\bra[1]{{\langle{#1}|}}
\newcommand\ket[1]{{|{#1}\rangle}}
\newcommand\ketbra[1]{\ket{#1}\!\bra{#1}}
\newcommand\braket[1]{\langle{#1}|{#1}\rangle}

\newcommand\density[1]{\calS({#1})}
\newtheorem{theorem}{Theorem}

\newtheorem{lemma}[theorem]{Lemma}

\title{
Additivity of Entanglement of Formation\\
of Two Three-level-antisymmetric States }
\author{Toshiyuki Shimono\\
\small{Department of Computer Science, University of Tokyo}\\
\small{7-3-1, Hongo, Bunkyo-ku, Tokyo, 113-0033, Japan}
}
\begin{document}\maketitle\thispagestyle{empty}
\begin{abstract}\INDENT
{\it Quantum entanglement} is the quantum information processing resource. 
Thus it is of importance to understand how much of entanglement 
particular quantum states have, and what kinds of laws  
entanglement and also transformation between entanglement states subject to. 
Therefore, it is essentialy important to use proper measures of entanglement 
which have nice properties. One of the major candidates of such measures is 
"entanglement of formation", 
and whether this measurement is additive or not is an important open problem. 
We aim at certain states so-called "antisymmetric states"
for which the additivity are not solved as far as we know,
and show the additivity for two of them.
%
\\

\noindent{\bf{}Keywords:}
quantum entanglement, entanglement of formation, additivity of entanglement measures,
antisymmetric states.
\end{abstract}%
\section{Introduction}   
\INDENT
Concerning the additivity of entanglement of formation,
 only a few results have been known. 
 Vidal et al. \cite{Vidal} showed that additivity holds for some mixture of Bell states and other examples 
by reducing the argument of additivity of the Holevo capacity of so-called "entanglement breaking quantum channels" %
\cite{Shor}
and they are the non-trivial first examples. 
 Matsumoto et al. \cite{Winter} showed that 
additivity of entanglement of formation holds for a family of mixed states 
by utilizing the additivity of Holevo capacity
for unital qubit channels \cite{King} 
via Stinespring dilation \cite{Stinespring}. 

%
%
%
%
%

In this extended abstract we prove 
that entanglement of formation is additive 
for tensor product of two three-dimensional bipartite 
antisymmetric states with a sketch of the proof.
We proved by combination of elaborate calculations.

\section{New additivity result}
\subsection{Antisymmetric states}\INDENT
Let us start with an introduction of our notations and concepts.
$\HH_-$ will stand for an antisymmetric Hilbert space, 
which is a subspace of 
a bipartite Hilbert space $ \HH_{AB}:=\HH_A\otimes\HH_B$,
where both $\HH_A$ and $\HH_B$ are $3-$dimensional Hilbert spaces,
spanned by basic vectors  $\{\ket{i}\}_{i=1}^3$.
$\HH_-$ is three-dimensional Hilbert space, spanned by states 
$\{\ket{i,j}\}_{ij=23,31,12}$,
where the state $\ket{i,j}$ is defined as $
\frac{\ket{i}\ket{j}-\ket{j}\ket{i}}  {\sqrt{2}}$.
%
%
 The space $\HH_-$ is called antisymmetric 
because by swapping the position of two qubits in any of its states $\ket{\psi}$
we get the state $-\ket{\psi}$. 
%
%
%
Let $\HH_-^\ox{n}$ be the tensor product of $n$ copies of $\HH_-$.
These copies will be discriminated by the upper index as
$\HH_-^{(j)}$, for $j=1\ldots n$. 
$\HH_-^{(j)}$ will then be an antisymmetric subspace 
 of $\HH_A^{(j)}\otimes\HH_B^{(j)}$.
%
\subsection{The result and proof sketch}
\INDENT

It has been shown in \cite{Vidal} that
 $E_f(\rho)=1$ for any mixed state $\rho\in\density{\HH_-}$. 
This result will play the key role in our proof.
We prove now that : 
\renewcommand{\thetheorem}{$\!\!$}
\begin{theorem}\label{th:1}
\begin{equation}\label{eq:1}
E_f(\rho_1\otimes\rho_2)= E_f(\rho_1) + E_f(\rho_2) \,\left(= 2\right) 
\end{equation}
for any ${\rho_1,\rho_2\in\density{\HH_-}}$. 
\end{theorem}
\begin{proof}
To prove this theorem, it is sufficient to show that 
\begin{equation}\label{eq:2}
 E_f(\rho_1\otimes\rho_2)\ge 2 
\end{equation}
since the subadditivity
 $E_f(\rho_1\otimes\rho_2)\le E_f(\rho_1)+ E_f(\rho_2) = 2$
  is trivial. 
{Indeed, it holds
\begin{eqnarray}
\!\!\!\!&E_f&\!\!\!\!(\rho_1\otimes\rho_2)=\inf\sum p_i E(\ketbra{\psi_i}) \nonumber\\
&\le&\!\!\!\! \inf\sum p_i^{(1)}p_i^{(2)} E(\ketbra{\psi_i^{(1)}}\otimes\ketbra{\psi_i^{(2)}})\nonumber\\
& = &\!\!\!\! \inf\sum p_i^{(1)} E(\ketbra{\psi_i^{(1)}})\nonumber\\
&&+\inf\sum p_i^{(2)} E(\ketbra{\psi_i^{(2)}})\nonumber\\
& = &\!\!\!\! E_f(\rho_1)+E_f(\rho_2)
\end{eqnarray} where 
$(p_i^{(j)},\ket{\psi_i^{(j)}})$ are subject to the condition of 
$\rho_j= \sum_i p_i^{(j)}\ketbra{\psi_i^{(j)}}$.
} 
To prove (\ref{eq:2}), we first show that
\begin{equation}\label{eq:3}
E(\ketbra{\psi})\ge 2, \text{ for any pure state }
\ignore{\forall} \ket{\psi}\in{\HH_-^\ox{2}}. 
\end{equation}	
Using the Schmidt decomposition, the state $\ket{\psi}$ can be decomposed
as follows:
\begin{equation}
\ket{\psi}=\sum\limits_{i=1}^3\sqrt{p_i}\:\ket{\psi_i^{(1)}}\otimes\ket{\psi_i^{(2)}},
\end{equation}
where $p_1,p_2,p_3>0,p_1+p_2+p_3=1$, and 
$\{\ket{\psi_i^{(j)}}\}_{i=1}^3$ is an orthonormal basis of 
the Hilbert space $\HH_-^{(j)}$, for $j=1,2$.
{
Note that this Schmidt decomposition is with respect to $\HH_-^{(1)} : \HH_-^{(2)} $,
or, it could be said that with respect to
 $\left(\HH_A^{(1)}\otimes\HH_B^{(1)}\right) : \left(\HH_A^{(2)}\otimes\HH_B^{(2)}\right)$, 
not with respect to $\left(\HH_A^{(1)}\otimes\HH_A^{(2)}\right) : \left(\HH_B^{(1)}\otimes\HH_B^{(2)}\right)$,
where ``:'' indicates how to separate the system into two subsystems for the decomposition.%
} 
\medskip

First, we will use the following fact.

\bigskip\noindent
\begin{lemma}\label{lem:1}
If $\{\ket{\psi_i}\}_{i=1}^3 $ is an orthonormal basis of $\HH_-$, 
then there exists an unitary operator $U$, acting on both $\HH_A$ and $\HH_B$, 
such that 
$U\otimes U $ maps the states $\ket{\psi_1},\ket{\psi_2},\ket{\psi_3}$ into
the states $ \ket{2,3},\ket{3,1},\ket{1,2}$, respectively.
\end{lemma}

\bigskip
Therefore, by this Lemma \ref{lem:1}, there exist unitary operators $U^{(1)},U^{(2)}$ such that
\begin{equation}
\begin{split}
\big( U^{(1)}&\otimes U^{(1)}\otimes U^{(2)}\otimes U^{(2)} \big)
\ket{\psi}\hfill\\
&=\sum\limits_{{i,j}\atop{ij=23,31,12}} 
 \sqrt{p_{ij}}\: \ket{i,j}\otimes\ket{i,j}
\;=:\;
 \ket{\psi'},
\end{split}
\end{equation}
where $p_{23}:=p_1,\; p_{31}:=p_2,\; p_{12}:=p_3$. 

\bigskip
As is written in the following, we use the following fact.

\medskip\noindent
\begin{lemma}\label{lem:2}
\begin{equation}
E(\ketbra{\psi'})\ge 2, \quad\text{ if }\quad
\begin{cases}
\quad p_{23},p_{31},p_{12}\ge0\\
\phantom{p_1}p_{23}+p_{31}+p_{12}=1
\end{cases}
.
\end{equation}
\end{lemma}
(We proved this lemma by solving a cubic equation and bounding the Shannon entropy function 
with polynomial functions.)
Local unitary operators do not change
von Neumann reduced entropy, 
and therefore $E ( \ketbra{\psi}) = E(\ketbra{\psi'})\ge 2$.
That is, the claim (\ref{eq:3}) is proven.

\medskip
We are now almost done. Indeed, the entanglement of formation is defined as 
\begin{equation}
E_f(\rho)
         =
\!\!  \inf_{  [(p_i,\psi_i)]_i \in{\Delta(\rho)}}  
 \sum_i p_i E(\ketbra{\psi_i})
\end{equation}
where 
$$\Delta(\rho)=   \left\{ \left[(p_i,\psi_i)\right]_i \Bigm|  \!\!
\begin{array}{l}
\sum_i p_i =1 , p_i >0 \forall i\\
\sum_i p_i \ketbra{\psi_i}=\rho, \braket{\psi_i}=1\forall i\!\!
\end{array}
\right\}  $$
and it is known that all $\ket{\psi_i} $ induced from $\Delta(\rho)$ satisfy 
$\ket{\psi_i}\in \Range(\rho)$, where $\Range(\rho)$ is sometimes called 
the image space of the matrix $\rho$, which is the set of $\rho\ket{\psi}$ with $\ket{\psi}$
running over the domain of $\rho$. Hence 
\begin{equation}
E_f(\rho)\ge \inf \left\{ E(\ketbra{\psi}) \bigm| \ket{\psi}\in\Range(\rho), \braket{\psi}=1 \right\} .
\end{equation}
Since $\rho_1\otimes\rho_2\in\calS(\HH_-^\ox{2})$, $\Range(\rho_1\otimes\rho_2)\subseteq \HH_-^\ox{2}$,
henceforth (\ref{eq:2}) is proven. 
Therefore  (\ref{eq:1}) have been shown.  \end{proof}
\section{Conclusions and discussion}
\INDENT
Additivity of the entanglement of formation for two three-dimensional bipartite antisymmetric states 
has been proven in this paper. 
The next goal could be to prove additivity for more than two antisymmetric states.
Perhaps the proof can utilize the value of lower bound of the reduced von Neumann entropy. 
Of course, the main goal is to show that entanglement of formation is additive, in general. 
However, this seems to be a very hard task.
%
%
\renewcommand{\thesection}{A.}

%
%
\newpage\onecolumn
{ 
\section{Appendix}
\INDENT 
We provide here proofs of two facts used in the proof of our main result.
\begin{lemma}
If $\{\ket{\psi_i}\}_{i=1}^3 \subset \HH_-$ is an orthonormal basis, 
there exists an unitary operator $U$, acting on both $\HH_A$ and $\HH_B$, 
such that 
$U\otimes U $ maps the states $\ket{\psi_1},\ket{\psi_2},\ket{\psi_3}$ into
the states $ \ket{2,3},\ket{3,1},\ket{1,2}$, respectively.
\end{lemma}
\begin{proof}
Let us start with notational conventions. In the following, 
${}^{T\!\!}{\Box}$ stands for the transpose of a matrix, 
${\phantom{}^{\phantom{T}\!\!} }{\Box}^*$ stands for taking complex conjugate of
each element of a matrix, 
${\phantom{}^{\phantom{T}\!\!} }{\Box}^\Theta$ denotes the transformation defined later.

\medskip
Let $U$ be represented as 
$\left(\begin{smallmatrix}u_{11}&u_{12}&u_{13}\\u_{21}&u_{22}&u_{23}\\u_{31}&u_{32}&u_{33}\end{smallmatrix}\right)$
with respect to the basis $\ket{1},\ket{2},\ket{3}$.%
For mathematicians, an operator and its matrix representation might be different 
objects, but for convenience, we identify 
$U$ with 
$\left(\begin{smallmatrix}
u_{11}&u_{12}&u_{13}\\u_{21}&u_{22}&u_{23}\\u_{31}&u_{32}&u_{33}\end{smallmatrix}\right)$
here. 
Lengthy calculations show that when a $9\times9$ dimensional matrix $U\otimes U$ is considered as mapping 
from $\HH_-$ into $\HH_-$, it can be represented by the following $3\times 3$ dimensional 
matrix, with respect to the basis $\ket{2,3},\ket{3,1},\ket{1,2}$,
 \vspace*{0mm}
\begin{equation*}
 U^{\Theta} :=\begin{pmatrix}
u_{22}u_{33}-u_{23}u_{32} && u_{23}u_{31}-u_{21}u_{33} && u_{21}u_{32}-u_{22}u_{31} \\
u_{32}u_{13}-u_{33}u_{12} && u_{33}u_{11}-u_{31}u_{13} && u_{31}u_{12}-u_{32}u_{11} \\
u_{12}u_{23}-u_{13}u_{22} && u_{13}u_{21}-u_{11}u_{23} && u_{11}u_{22}-u_{12}u_{21} 
\end{pmatrix} . \vspace*{0mm}
\end{equation*} 
One can then show that\vspace*{0mm}
$$U^{\Theta}\cdot ^{\:T\!\!\!\!}U = (\det U)
\left(\begin{smallmatrix}1&0&0\\0&1&0\\0&0&1\end{smallmatrix}\right),\vspace*{0mm}
$$
and by multiplying with $U^*$ from the right in the above equation,
 one obtain 
 $U^\Theta = (\det U)\cdot U^*$,
since $U$ is an unitary matrix, and ${}^{T\!\!\!}U \cdot U^*$ is equal to the identity matrix.
\\

Since  $\{\ket{\psi_i}\}_{i=1,2,3}  $
is an orthonormal basis of $\HH_-$,
there exists an unitary operator on $\HH_-$
such that 
$\ket{\psi_1}\mapsto\ket{2,3}, \ket{\psi_2}\mapsto\ket{3,1}, \ket{\psi_3} \mapsto\ket{1,2}$,
and let $\Theta_\psi$ be the corresponding matrix with respect to the basis
$ \{\ket{i,j}\}_{ij=23,31,12}$.

Let 
 $ U_\psi:={({\det \Theta_\psi})^\frac{1}{2}}\cdot{\Theta_\psi^*}$ 
.\footnote{In the above definition it does not matter which of two roots of 
$\det \Theta_\psi$ are taken}
It holds  $U_\psi^\Theta=\Theta_\psi$.
\footnote{
 Indeed,
$ U_\psi^\Theta  =(\det U_\psi){ U_\psi^*}
= (\det \Theta_\psi)^{\frac{3}{2}} \det \Theta_\psi^*
\cdot ({(\det \Theta_\psi)^{\frac{1}{2}} })^*
\Theta_\psi 
= \Theta_\psi $ .
Note that $ \det U_\psi
= \det( \det(\Theta_\psi)^{\frac{1}{2}}\,  \Theta_\psi^* )
=(\det \Theta_\psi)^{\frac{3}{2}} \det \Theta_\psi^* $
because $\Theta_\psi^*$ is a $3\times 3 $ matrix.
} Therefore $U_\psi\otimes U_\psi = U'_\psi$. 
The operator $U_\psi$ is the one needed to satisfy the statement of Lemma \ref{lem:1}.
\qed\end{proof}

\begin{lemma}
\begin{equation*}
E(\ketbra{\psi'})\ge 2 \quad\text{ if }\quad
\left\{
\begin{array}{l}
\ket{\psi'} = \sum\limits^{i,j}_{ij=23,31,12}\sqrt{p_{ij}}\:\ket{i,j}\ket{i,j}  \\
 p_{23},p_{31},p_{12}\ge0\\
p_{23}+p_{31}+p_{12}=1
\end{array}\right.
.
\end{equation*}
\end{lemma}
\begin{proof}
Let $p_{32}:=p_{23} , p_{13}:=p_{31} , p_{21}:=p_{12}$.
Then it holds, \vspace*{0mm}
\begin{eqnarray*}
\ket{\psi'}  
&=& \sum^{i,j}_{1\le i < j\le 3}\sqrt{p_{ij}}\:\ket{i,j}\ket{i,j}  \vspace*{-6mm}
										\\
&=&\frac{1}{{2}}\sum^{i,j}_{1\le i < j\le 3}\sqrt{p_{ij}}\:     
\{\ket{ii;jj}-\ket{ij;ji}-\ket{ji;ij}+\ket{jj;ii}\}
                                                                                \vspace*{-6mm} \\
&=&\frac{1}{{2}}\sum^{i,j}_{1\le i\ne j\le 3}\sqrt{p_{ij}}\: 
\{\ket{ii;jj}-\ket{ij;ji}\} ,
                                                                       \vspace*{0mm}
\end{eqnarray*}                                                                                 
 where $\ket{i_1i_2;i_3i_4}$ denotes the tensor product 
$\ket{i_1}\otimes\ket{i_2}\otimes\ket{i_3}\otimes\ket{i_4}$ , 
 $\ket{i_1}\in\HH_A^{(1)}$, $\ket{i_2}\in\HH_A^{(2)}$, $\ket{i_3}\in\HH_B^{(1)}$  and  $\ket{i_4}\in\HH_B^{(2)}$ 
,
and the condition $1\le i\ne j\le 3$ actually means 
"$1\le i\le 3$ and $1\le j\le 3$ and   $i\ne j$". 
This convention will be used also in the following. 

We are now going to calculate the reduced matrix of $\ketbra{\psi'}$, 
which we will denote as $\Xi$, and it will be 
decomposed into the 
direct sum as follows. 
\begin{eqnarray}
 \Xi &:=& \mathop{\mathrm{Tr}}\limits_{\HH_B^{(1)}\otimes\HH_B^{(2)} } \ket{\psi'}\bra{\psi'}  
 \nonumber          \\
 &=& \frac{1}{4}\sum^{i,j,k,l}_{{1\le i\ne j\le 3}\atop{1\le k\ne l\le 3}}
\sqrt{p_{ij}p_{kl}}\mathop{\mathrm{Tr}}\limits_{\HH_B^{(1)}\otimes\HH_B^{(2)} }
\left(
{
  {\ket{ii;jj}\bra{kk;ll}-\ket{ii;jj}\bra{kl;lk}\phantom{abc}}
  \atop%
   {\phantom{abc}-\ket{ij;ji}\bra{kk;ll}+\ket{ij;ji}\bra{kl;lk} }
}
\right)
 \nonumber          \\
 &=& \frac{1}{4}\sum^{i,j,k,l}_{{1\le i\ne j\le 3}\atop{1\le k\ne l\le 3}}
\sqrt{p_{ij}p_{kl}}\mathop{\mathrm{Tr}}\limits_{\HH_B^{(1)}\otimes\HH_B^{(2)} }
\big(
{
  {\ket{ii;jj}\bra{kk;ll}+\ket{ij;ji}\bra{kl;lk} }
}
\big)
 \nonumber          \\
 &=&
 \frac{1}{4}\sum^{i,j,k}_{{1\le i\ne k\le 3}\atop{1\le j\ne k\le 3}}
 \sqrt{p_{ik}p_{jk}}\: \ket{ii}\bra{jj}
+
 \frac{1}{4}\sum^{i,j}_{{1\le i\ne j\le 3}}
 {p_{ij}}\: \ket{ij}\bra{ij}   
                       \nonumber          \\
   &   \cong &
\frac{1}{4}
\left ( 
\begin{smallmatrix} 
 p_{12}+p_{13} & \sqrt{p_{13}p_{23}} & \sqrt{p_{12}p_{23}} \\
 \sqrt{p_{13}p_{23}} & p_{12}+p_{23} & \sqrt{p_{12}p_{13}} \\
 \sqrt{p_{12}p_{23}}& \sqrt{p_{12}p_{13}} &p_{13}+p_{23}
\end{smallmatrix}
\right)  
\oplus
\frac{1}{4}(p_{12}) ^{\oplus 2} \oplus
\frac{1}{4}(p_{13}) ^{\oplus 2} \oplus
\frac{1}{4}(p_{23}) ^{\oplus 2} ,                       \label{eq:direct_sum}
\end{eqnarray}
where $\oplus$ denotes the direct sum of matrices,
and $\Box^{\oplus n}$ denotes the direct sum of $n$ copies of the same matrix.

\medskip
We need to get eigenvalues of $\Xi$ 
in order to calculate reduced von Neumann entropy 
$$
E(\ketbra{\psi'})
 = -\Tr\left(\Xi\log_2\Xi\right)
 = -\!\!\sum\limits_{\lambda : \mathrm{e.v. of }\;\Xi}\!\!\lambda\,\log_2\lambda.$$
In this case, fortunately, the eigenvalues can be determined explicitly
from the expression (\ref{eq:direct_sum}). 
They are the following ones: 
\begin{equation}
\left(
\frac{1-\cos\theta}{6}, 
\frac{1-\cos(\theta+\frac{2\pi}{3})}{6}, 
\frac{1-\cos(\theta+\frac{4\pi}{3})}{6}, 
\frac{p_{12}}{4},\frac{p_{12}}{4},
\frac{p_{13}}{4},\frac{p_{13}}{4},
\frac{p_{23}}{4},\frac{p_{23}}{4}
\right)
\label{ex:explicit}
\end{equation}
 for a certain $-\frac{\pi}{3}<\theta\le\frac{\pi}{3}$.%
\footnote{
The exact value of $\theta$ will be no importance for us.} 
These eigenvalues are denoted as 
$
(\lambda_1, \lambda_2, \ldots
,\lambda_9),
$ respectively. Although $\lambda_4,\dots,\lambda_9$ are trivial, 
$\lambda_1,\lambda_2,\lambda_3$ are the roots of the cubic polynomial
\begin{equation}\label{eq:theequation}
g(\lambda) := \lambda^3-\frac{1}{2}\lambda^2+\frac{1}{16}\lambda-\frac{p_{12}\:p_{13}\:p_{23}}{16}
\end{equation}
that is the characteristic polynomial function of the cubic matrix that appeared in the expression
(\ref{eq:direct_sum}).
%
%
We must solve this cubic equation to obtain (\ref{ex:explicit}).
The cubic equation $g(\lambda)=0$ is in Cardan's irreducible form,%
\footnote{A cubic equation is said to be in Cardan's irreducible form if 
its three roots are real. }
 because $\Xi$ is the density matrix.
 In such a case, the roots of the cubic equation are \vspace*{0mm}
\begin{equation}\label{eq:tri}
\alpha+\beta\cos\theta,
\alpha+\beta\cos(\theta+\frac{2\pi}{3}),
\alpha+\beta\cos(\theta+\frac{4\pi}{3}).\vspace*{0mm}
\end{equation}
One can easily show that
 $ \lambda_1+\lambda_2+\lambda_3=3\alpha,$ and $ 
\lambda_1^2+\lambda_2^2+\lambda_3^2=3\alpha^2+\frac{3}{2}\beta^2 $.
If $\lambda_1,\lambda_2,\lambda_3$ are equal to the roots of 
the cubic equation $\lambda^3+a_1\lambda^2+a_2\lambda+a_3=0$, then
$\lambda_1+\lambda_2+\lambda_3=-a_1$,
$\lambda_1^2+\lambda_2^2+\lambda_3^2=a_1^2-2a_2 $.
Taking $a_1=-\frac{1}{2},a_2=\frac{1}{16}$ from the expression (\ref{eq:theequation}),
we get the following system of equations $3\alpha=\frac{1}{2},
3\alpha^2+\frac{3}{2}\beta^2 = \frac{1}{8} $,
and $(\alpha,\beta)=\left(\frac{1}{6},-\frac{1}{6}\right)$ is sufficient.
Applying this argument into (\ref{eq:tri}), we complete (\ref{ex:explicit}).

\bigskip  
Our idea is now to show that \vspace*{0mm}
\begin{equation}\label{eq:00}
E(\ketbra{\psi'}) = \sum_{i=1}^9(-\lambda_i\log_2\lambda_i)\ge 2 . 
\vspace*{0mm}
\end{equation}
 This will be shown if we prove that it holds \vspace*{0mm}
\begin{equation}\label{eq:01}
\sum_{i=1}^3(-\lambda_i\log_2\lambda_i)\ge 1 \text{ and } 
\sum_{i=4}^9(-\lambda_i\log_2\lambda_i)\ge 1.\vspace{0mm}
\end{equation}
The second inequalities is easy to verify by simple calculations.
To finish the proof 
of the lemma
we therefore need to show that 
\begin{equation}\label{eq:last}
\sum_{i=1}^3(-\lambda_i\log_2\lambda_i)\ge 1.
\end{equation} 

\medskip
Without loss of generality, we assume $\theta\in\left[0,\frac{\pi}{3}\right]$.%
\footnote{
The sequence of $\{\lambda_i\}_{i=1}^3$ doesn't change if $\theta$ is 
replaced by $-\theta$. Thus we can change the assumption 
$\theta\in\left(-\frac{\pi}{3},\frac{\pi}{3}\right]$, into
$\theta\in\left[0,\frac{\pi}{3}\right]$. } 
Clearly,
$\lambda_1\in \left[0, \frac{1}{12}\right]$ and 
$
\lambda_2,\lambda_3=\frac{1}{4} -\frac{\lambda_1\pm\sqrt{\lambda_1-3\lambda_1^2}}{2}
 \in\left[\frac{1}{12}, \frac{1}{3} \right]
$ 
($\lambda_2,\lambda_3$ can be regarded as the solution of the following systems of equations: 
$\lambda_1+\lambda_2+\lambda_3=\frac{1}{2},
\lambda_1^2+\lambda_2^2+\lambda_3^2=\frac{1}{8}$ ) .
You can also show that 
\begin{equation}\label{eq:2ineq}-z \log_2 z 
\ge
\begin{cases}
\phantom{ab}( \log_2 12)\:z 
  & 
\text{ if $ z\in\left[0,\frac{1}{12}\right]$} 
     \\
\phantom{ab} \frac{1}{2}+\frac{\log_{e} 4 -1}{\log_{e} 2}(z-\frac{1}{4})-4(z-\frac{1}{4})^2
  & 
\text{ if $z\in\left[\frac{1}{12},\frac{1}{3}\right]$}
\end{cases}
\end{equation}
(see Fig.1).
\begin{figure}\label{fig:1}
\begin{center}\includegraphics[width=0.75\linewidth]{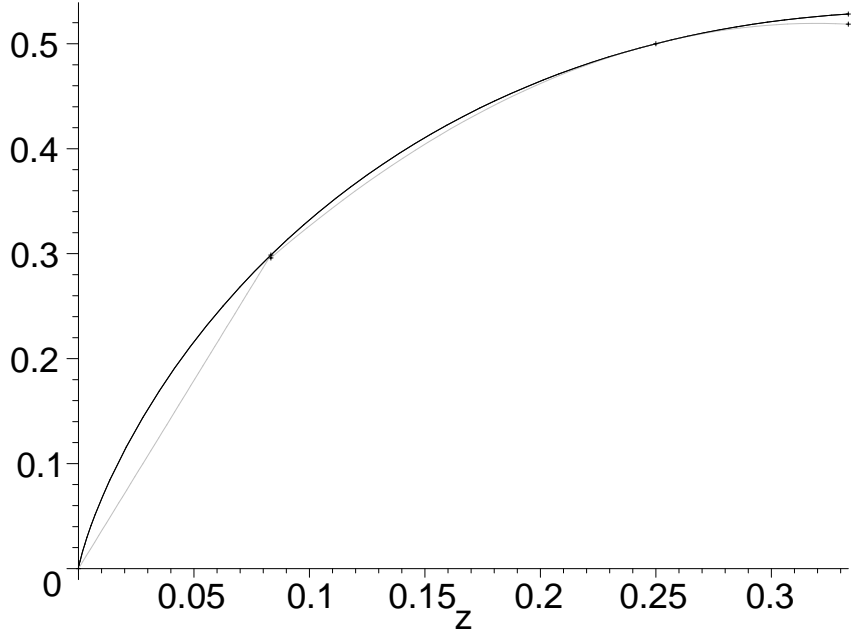}\end{center}
\caption{By virture of Shannon entropy function being lower-bounded by the polynomial functions,
we can evaluate the von Neumann entropy of $\Xi$.
}
\end{figure}
The first inequality of (\ref{eq:2ineq}) is easily confirmed.
On the other hand, one way of the proof of the second inequality is as follows: 
Let  $
f(z):=\bigl(-z\log_2 z\bigr) - 
\left(\frac{1}{2}+\frac{\log_{e} 4 -1}{\log_{e} 2}(z-\frac{1}{4})
-4(z-\frac{1}{4})^2\right)$.
Differentiating this expression by $z$ once and twice, we can get 
the increasing and decreasing table as follows. 
\begin{center}
\begin{tabular}{|c||c|c|c|c|c|c|c|} 
\hline
 $z$     & $\frac{1}{12}$ &    & $\frac{1}{8\log_{e} 2} $ &   &  $\frac{1}{4}$ &  
& $\frac{1}{3}$ \\ \hline\hline
$f(z)$ &   $+$  & $\curvearrowright$  &$+$ & $\searrow$ &$0$ &$\nearrow$ &  \\ \hline
$f'(z)$  &        &    &             & $-$  & $ 0 $  & $+$ &       \\ \hline
$f''(z)$ &        & $-$&     $0 $    & $+$  & $ + $  & $+$ &       \\ \hline
\end{tabular} 
\end{center}
The table indicates 
 $ f(z)\ge 0 $  for $ z\in\left[\frac{1}{12},\frac{1}{3}\right]$. 
Now we indeed get the lower bounds by polynomial functions. 

Combining all of the above inequalities
we get 
(\ref{eq:last}) as
\begin{equation*}
-\sum\limits_{i=1}^3 \lambda_i \log_2 \lambda_i \ge 1 + 
\left(\frac{\log_{e} 3 +2}{\log_{e} 2}-2\right)\lambda_1  + 4\lambda_1^2 \ge 1 \quad .
\end{equation*}
so that (\ref{eq:01}) and (\ref{eq:00}) are successively shown and that our proof is finished.
\end{proof}
}%

\begin{thebibliography}{3}
\bibitem{Vidal}G. Vidal, W. D\"ur, J.~I. Cirac, Entanglement Cost of Bipartite Mixed States (2002),
Physical Review Letters, {\bf 89}, 027901.
\bibitem{Shor}Peter W. Shor (2002), Additivity of the Classical Capacity of 
Entanglement-Breaking Quantum Channels,
quant-ph/0201149.
\bibitem{Winter}K. Matsumoto, A. Winter, T. Shimono (2002), 
Remarks on additivity of the Holevo channel capacity and of the entanglement of formation, 
quant-ph/0206148.
\bibitem{King}Christopher King, Additivity for unital qubit channels (2001),
quant-ph/0103156.
\bibitem{Stinespring}W. F. Stinespring (1955), Positive functions on $C^*$--algebras, 
Proceedings of the American Mathematical Society,
{\bf 6}, 211-216.
\end{thebibliography}
\end{document}